\documentclass[aps,showkeys,preprint]{revtex4}
\usepackage{amsmath,amssymb,amsthm}

\usepackage{amsmath}
\usepackage{graphicx}

\newcommand{\beq}{\begin{equation}}
\newcommand{\eeq}{\end{equation}}
\newcommand{\Au}{{A_\uparrow}}
\newcommand{\Ad}{{A_\downarrow}}

\newcommand{\tu}{{\tau_\uparrow}}
\newcommand{\td}{{\tau_\downarrow}}

\newcommand{\cA}{{\cal A}}
\newcommand{\cL}{{\cal L}}
\newcommand{\cS}{{\cal S}}

\newcommand{\cLu}{{\cal L}^{\uparrow}}
\newcommand{\cLd}{{\cal L}^{\downarrow}}
\newcommand{\cK}{{\cal K}}
\newcommand{\cM}{{\cal M}}
\newcommand{\cP}{{\cal P}}

\newcommand{\cMu}{{\cal  M}^{\uparrow}}
\newcommand{\cMd}{{\cal M}^{\downarrow}}
\newcommand{\Li}{{\text {Li}}}

\newcommand{\cC}{{\cal C}}
\newcommand{\phiu}{\phi^{\uparrow}}
\newcommand{\phid}{\phi^{\downarrow}}
\newcommand{\wu}{w^{\uparrow}}
\newcommand{\wdd}{w^{\downarrow}}

\newcommand{\norm}[1]{\left \|{#1}\right \|}
\newcommand{\abs}[1]{\left | {#1}\right | }
\newcommand{\all}[2]{ \left \{\, {#1} \, : \, {#2} \, \right \} }

\newcommand{\hip}{H^\infty(\Pi)}

\newtheorem{theorem}{Theorem}
\newtheorem{prop}[theorem]{Proposition}
\newtheorem{lem}[theorem]{Lemma}

\newtheorem{corollary}[theorem]{Corollary}

\begin{document}

\title{Asymptotics of the Farey Fraction Spin Chain Free Energy at the Critical Point}

\author{Oscar F. Bandtlow}
\email{o.bandtlow@qmul.ac.uk}
\affiliation{School of Mathematical Sciences, 
Queen Mary University of London,
Mile End Road, London E1 4NS, United Kingdom\,}
\author{Jan Fiala}
\email{jfiala@my.lamar.edu}
\affiliation{Department of Chemistry and Physics, Lamar University,
P.O. Box 10046, Beaumont, TX 77710, USA}
\author{Peter Kleban}
\email{kleban@maine.edu}
\affiliation{LASST and Department of Physics \& Astronomy,
University of Maine, Orono, ME 04469, USA}
\author{Thomas Prellberg}
\email{t.prellberg@qmul.ac.uk} 
\affiliation{School of Mathematical Sciences, 
Queen Mary University of London,
Mile End Road, London E1 4NS, United Kingdom}

\date{\today}
\begin{abstract} 
We consider the Farey fraction spin chain in an external field $h$.  Using 
ideas from dynamical systems and functional analysis, we show that the free 
energy $f$ in the vicinity of the second-order phase transition is given, exactly,
by 
$$
f \sim \frac t{\log t}-\frac1{2} \frac{h^2}t \quad \text{for} \quad h^2\ll t 
\ll 1 \;.
$$
 Here $t=\lambda_{G}\log(2)(1-\frac{\beta}{\beta_c})$ is a reduced temperature, so 
that the deviation from the critical point is scaled by the 
Lyapunov exponent of the Gauss map, $\lambda_G$.  It follows that $\lambda_G$ 
determines the amplitude of both the specific heat and susceptibility 
singularities.  To our knowledge, there is only one other  microscopically 
defined interacting model for which the free energy near a phase transition is known 
as a function of two variables.

 Our results confirm what was found previously with a cluster 
approximation, and show that a clustering mechanism is in fact responsible for the 
transition. However, the results disagree in part with a renormalisation group 
treatment.
\end{abstract}
\maketitle

\section{Introduction}

The theory of second-order phase transitions has a long and well-developed 
history.  However, for spin models coupled to an external magnetic field 
there are very few exact microscopic calculations for the 
free energy $f(\beta,h)$ as a function of both inverse temperature $\beta$ {\it and} 
magnetic field strength
$h$ in the vicinity of such a transition, and (to our knowledge) with the exception of 
the ice-rule models (see Section V. E. of \cite{LW}), these are limited to mean-field, 
non-interacting models (e.g. the spherical and Kac-van der Waals models \cite{BK}), 
or other similarly defined models \cite{FF}. 

In this work, by using operator techniques, we calculate the exact free energy $f(\beta,h)$ for a 
model with many-body long-range interactions \cite{KO}. This model has been investigated
previously with a cluster approximation \cite{PFK}, replacing the non-trivial many-body 
interactions by interactions within clusters, and thereby leading to a model very similar to \cite{FF}. 
Our present work confirms the validity of that cluster approximation near the critical point.
Intriguingly, we also find that both critical amplitudes in the free energy scale with 
a Lyapunov exponent.

Phase transitions in one-dimensional systems are unusual, essentially because, 
as long as the interactions are of finite range and strength, any putative 
ordered state at finite temperature will be disrupted by thermally induced 
defects, and a defect in one dimension is very effective at destroying order. 
On the other hand, long range or infinite interactions generally make the model 
ordered at all finite temperatures. Despite this, there are a number of 
examples of one-dimensional systems that exhibit a phase transition. The Farey 
fraction spin chain  \cite{KO} is one such case, which has attracted interest 
from both physicists and mathematicians (see \cite{FK2,KOPS,Pe,B,CK} and references therein).  
This model has a phase transition at a finite temperature. While the transition is 
of second-order, it has some properties that are usually found with a first-order transition: 
for external field $h=0$, the magnetisation jumps from completely saturated, below the 
transition, to zero above it.  Despite this unusual behaviour, the model does not violate 
scaling theory, but rather is encompassed as a limiting case \cite{FK}.  

In some recent work, \cite{FK,PFK}, this model has  been generalised to finite 
external field, and analysed via both renormalisation group methods and with a 
dynamical system-inspired cluster approximation.  Neither method is rigorous, 
and the results are not quite the same.  Specifically, the dependence of 
$f(\beta,h)$ on $h$ differs by logarithmic factors.  This motivates a 
more rigorous analysis of the model.  We find that the cluster picture indeed leads 
to the correct result for the asymptotic form for $f(\beta,h)$, and, in 
addition, we are able to evaluate the constants.  Intriguingly, they involve
$\lambda_G$, the Lyapunov exponent of the Gauss map. This arises naturally here, 
since the Gauss map is intimately related to the first-return map 
of the Farey map, which specifies the transfer operator giving the Farey fraction 
spin chain partition function \cite{FK2}.

In Section \ref{sec:model} we define the model, first in the standard way using 
matrices, and then via transfer operators.   
In Section \ref{sec2} we  derive some operator identities that are necessary 
for our analysis, and in Section \ref{sect_funct} we 
specify a function space and study the spectral properties of the relevant transfer
operators. Section \ref{sect_dyn} describes the
connection to the Gauss map. Section \ref{sec3} is the heart of 
our work.  Here we use perturbation theory around the critical point $(\beta,h)=(1,0)$ to find 
the asymptotic behaviour of the free energy $f(\beta,h)$.  The key to our method 
is the use of a ``cluster operator'', which encodes the behaviour of clusters of 
up and down spins while possessing tractable spectral properties, thus validating the results
obtained with the cluster approximation of \cite{PFK}.  

\section{The Model}
\label{sec:model}

The Farey fraction spin chain may be constructed, for inverse temperature 
$\beta$ and magnetic field $h$, via the two matrices 
\beq
\Au=\begin{pmatrix}1&0\\1&1\end{pmatrix}
\quad\text{and}\quad
\Ad=\begin{pmatrix}1&1\\0&1\end{pmatrix}
\; ,
\eeq
which correspond to ``spin up" and ``spin down", respectively.
The spin chain partition function comes in various guises, all of which have 
the same free energy (at least for $h=0$, see \cite{KO,FKO,FK,FK2}).  
Here we are considering the generalised Knauf spin chain \cite{FK2}, not the 
``trace'' model studied in \cite{PFK}.  We make this choice for technical reasons.  
However, by universality, our results 
are supposed to apply to any of the Farey spin chains (see  
\cite{KO,FKO,FK,FK2} for definitions of the various chains). 

Defining matrix products 
\begin{equation}\label{M}
M_N:=\prod_{i=1}^{N}\Au^{1-\sigma_i}\Ad^{\sigma_i},\qquad \sigma_i\in \{0,1\},
\end{equation}
where the dependence of $M_N$ on the $\sigma_i$ has been suppressed,
and writing a given matrix product as $M_N =
\bigl( \begin{smallmatrix}
a&b\\ c&d
\end{smallmatrix} \bigr)
$ 
we define the spin chain partition function by
\beq \label{Zdef}
Z_N(\beta, h;x)=\sum_{\{\sigma_i\}}\frac1{(cx+d)^{2\beta}}e^{ -\beta h\left ( 
2\sum_{i=1}^{N}\sigma_i - N  \right )},
\end{equation}
where $x\geq0$ is a parameter (which does not affect the free energy, as we shall argue below).
When $M_N$ begins with $\Au$, $c$ and $d$ are neighbouring Farey denominators at 
level $N$ in the modified Farey sequence (see \cite{KO} for further details on 
this connection), whence the nomenclature ``Farey" for this spin chain model.

The free energy $f(\beta,h)$ is defined via
\beq
-\beta f(\beta,h) = \lim_{N \to \infty} \frac1N\log Z_N(\beta, h;x)\;.
\eeq

Alternatively, the partition function can be expressed using transfer 
operators.  In
order to emphasise the difference between matrices and operators, we denote the 
latter with script letters. We begin by defining the operator 
\beq \label{xfr}
\cL_{\beta,h}=e^{-\beta h}\cLu_\beta+e^{\beta h}\cLd_\beta
\eeq
where
\beq
\cLu_\beta f(x)
=\frac1{(1+x)^{2\beta}}f\left(\frac x{1+x}\right)
\quad\text{and}\quad
\cLd_\beta f(x)
=f(1+x)\;.
\eeq
Thus we obtain, as in \cite{FK2} 
\beq
Z_N(\beta,h;x)=\cL_{\beta,h}^N1(x)\;.
\eeq
This expression indicates that the free energy $f(\beta,h)$ is given by the logarithm 
of the spectral radius of $\cL_{\beta,h}$ on a suitable function space,
\begin{equation}\label{frad}
-\beta f(\beta,h;x) =\log r(\cL_{\beta,h})\;.
\end{equation}
We will return to this point below when we specify the function space used in our analysis.

There is another notation that is sometimes used in the literature, which we mention for
completeness and for comparison with previous work on this model. Using a ``slash'' notation which is standard in number theory,
the action of a $2\times2$ matrix [$M 
 = 
 \bigl( \begin{smallmatrix}
a&b\\ c&d
\end{smallmatrix} \bigr)
 \in SL_2(\mathbb Z)$]
on a function $f$ is defined via 
\beq \label{slash}
f(x)\left|\begin{pmatrix}a&b\\c&d\end{pmatrix}\right.=\frac1{(cx+d)^{2\beta}}\,f
\left(\frac{ax+b}{cx+d}\right)\;.
\eeq
Thus (\ref{Zdef}) can be written as
\beq \label{pf}
Z_N(\beta,h;x)=1(x)\left|(e^{-\beta h}\Au+e^{\beta h}\Ad)^N\right.\;.
\eeq
Note that in order to be consistent with the group structure of $SL_2(\mathbb Z)$, any 
addition and scalar multiplication is performed {\it after} the matrix action 
on the function has been computed.

In the disordered (high-temperature) phase, we expect that there is a leading 
eigenvalue $\lambda (\beta,h)$  of ${\cal L}_{\beta,h}$ which satisfies 
$\lambda (\beta,h)>1$,  is non-degenerate, and belongs to the discrete spectrum,
and that the free energy is given by
\begin{equation}\label{flam}
-\beta f(\beta,h;x) =\log \lambda(\beta,h)\; ,
\end{equation}
which is independent of $x$.

There is an obvious (spin flip) symmetry in our model. Since 
\beq
S\Au S=\Ad\quad\text{with}\quad S=\begin{pmatrix}0&1\\1&0\end{pmatrix}\;,
\eeq
it is natural to define the corresponding operator $\cS_\beta$ as
\beq
\cS_\beta f(x)
=x^{-2\beta}f(1/x)\;.
\eeq
Note that $S^{-1}=S$ and $\cS_\beta^{-1}=\cS_\beta$, i.~e.~both $S$ and 
$\cS_\beta$ are involutions. For the transfer operators we find
\beq \label{sconj}
\cLu_\beta=\cS_\beta\cLd_\beta\cS_\beta\quad\text{and}\quad\cL_{\beta,h}=
\cS_\beta\cL_{\beta,-h}\cS_\beta\;.
\eeq

\begin{figure}[ht]
\centering\includegraphics[width=10.0cm]{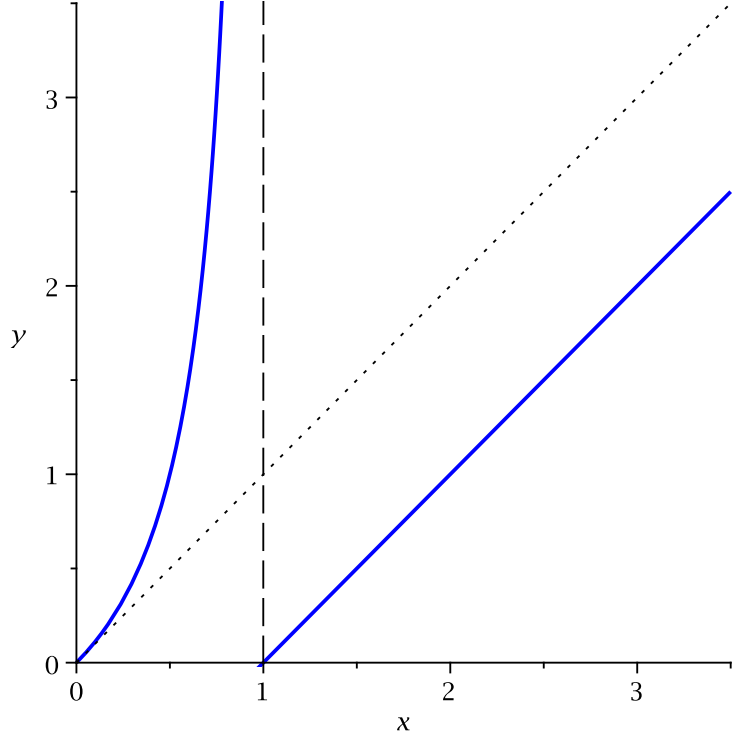}
\caption{The graph of the map $y=T(x)$.} \label{fig1}
\end{figure}

The operator $\cL_{\beta,0}$ has a nice interpretation as the 
Ruelle-Perron-Frobenius transfer operator \cite{beck93} for the dynamical system given by 
iteration of the map
\beq
\label{trafo}
T(x)=\left\{\begin{array}{ll}
x/(1-x)\;,&\quad0\leq x<1\\
x-1\;,&\quad x\geq1 \;,
\end{array}\right.
\eeq
and $\cL_{\beta,h}$ can be viewed as a weighted generalisation.
Note that the map $T$ has the symmetry $T(1/x)=1/T(x)$. This map differs from 
the Farey map used in, say, \cite{FK2}. The graph of this map is shown in
Fig. \ref{fig1}. One clearly observes the existence of two marginal neutral
fixed points, at the origin and infinity.

Now consider the generating function
\beq
G(\beta,h,z;x)=\sum_{N=0}^\infty z^N Z_N(\beta,h;x)\;.
\eeq
Examination of $G$ motivates the operator relations discussed below and makes 
a connection with the treatment in \cite{PFK}. 
One may rewrite $G$ in terms of the resolvent $[1-z\cL_{\beta,h}]^{-1}$
of $\cL_{\beta,h}$ as
\begin{eqnarray}
\label{eq:res}
G(\beta,h,z;x)
              &=&[1-z\cL_{\beta,h}]^{-1}1(x)\;.
\end{eqnarray}
Equation (\ref{eq:res}) indicates that $z_c(\beta,h)$ 
is  equal to the inverse of the spectral radius $1/r(\cL_{\beta,h})$.
The free energy is then given as
\begin{equation}\label{lfe}
\beta f(\beta,h)=\log z_c(\beta,h)\;, 
\end{equation}
where $z_c(\beta,h)$ is the smallest singularity of $G(\beta,h,z;x)$ in $z$ on 
the positive real axis. 
Thus, in principle, we could find the free energy by analysing $G$.  
However, it is very difficult to do this directly, since $\cL_{\beta,h}$ is 
not sufficiently well-behaved. In order to make progress we resort below
to a more nuanced treatment.

\section{Identities and Spectral Relations} \label{sec2}

This section introduces a Lemma that is the basis of our 
analysis.  It allows us to avoid dealing directly with $\cL_{\beta,h}$, which 
is difficult to control at the critical point $(\beta,h)=(1,0)$.

To motivate this section, let us consider an arrangement of $N$ spins. We can
collect adjacent spins of equal orientation into clusters of the form
$$
\underbrace{\uparrow\cdots\uparrow}_{\geq0}\;
\overbrace{\underbrace{\downarrow\cdots\downarrow}_{\geq1}\;
\uparrow
\cdots
\downarrow\;
\underbrace{\uparrow\cdots\uparrow}_{\geq1}
}^{\mbox{$n$ pairs, $n\geq0$}}\;
\underbrace{\downarrow\cdots\downarrow}_{\geq0}\;.
$$
Notice that configurations starting and ending with either spin orientation are 
included. Such an arrangement of spins corresponds uniquely to a particular product
of operators $\cLu_\beta$ and $\cLd_\beta$,
$$
\underbrace{\cLu_\beta\cdots\cLu_\beta}_{\geq0}\;
\overbrace{\underbrace{\cLd_\beta\cdots\cLd_\beta}_{\geq1}\;
\cLu_\beta
\cdots
\cLd_\beta\;
\underbrace{\cLu_\beta\cdots\cLu_\beta}_{\geq1}
}^{\mbox{$n$ pairs, $n\geq0$}}\;
\underbrace{\cLd_\beta\cdots\cLd_\beta}_{\geq0}\;,
$$
and taking a weighted sum over all possible spin configurations of arbitrary length,
we find
\begin{multline}
[1-z\cL_{\beta,h}]^{-1}=[1-ze^{-\beta h}\cLu_\beta-ze^{\beta h}\cLd_\beta]^{-1}=\\
[1-ze^{-\beta h}\cLu_\beta]^{-1}
\sum_{n=0}^\infty\left(
ze^{\beta h}\cLd_\beta[1-ze^{\beta h}\cLd_\beta]^{-1}
ze^{-\beta h}\cLu_\beta[1-ze^{-\beta h}\cLu_\beta]^{-1}
\right)^n
[1-ze^{\beta h}\cLd_\beta]^{-1}\;.
\end{multline}
We now introduce the operators
\beq\label{clusters}
\cMu_{\beta,\tau}=\tau \cLu_\beta[1-\tau \cLu_\beta]^{-1}
\quad\text{and}\quad
\cMd_{\beta,\tau}=\tau \cLd_\beta[1-\tau \cLd_\beta]^{-1} \; .
\eeq
Notice that as a formal power series in $\tau$,
\beq
 \cMu_{\beta,\tau}=\sum_{n=1}^\infty \tau^n{\cLu_\beta}^n
\quad\text{and}\quad
\cMd_{\beta,\tau}=\sum_{n=1}^\infty \tau^n{\cLd_\beta}^n\;.
\eeq
To keep the discussion general, we shall assume in the following that
$\cLu_\beta$ and $\cLd_\beta$ are bounded operators on a Banach space,
which will be specified later.

Then $\cMu_{\beta,\tau}$ and $\cMd_{\beta,\tau}$ exist whenever the resolvents $[1-\tau \cLu_\beta]^{-1}$ and 
$[1-\tau \cLd_\beta]^{-1}$ exist, i.e.
for $\tau^{-1} \notin \sigma(\cLu_{\beta})$ or 
$\tau\notin\sigma(\cLd_{\beta})$, respectively (here $\sigma({\cal A})$ denotes 
the spectrum of the operator ${\cal A}$). 
This motivates the following identities.

\begin{lem}\label{lem1}
Let $z^{-1}\notin\left(\sigma(e^{-\beta h}\cLu_{\beta})\cup\sigma(e^{\beta h}\cLd_{\beta})\right)$. Then
\beq \label{oid}
[1+ \cMd_{\beta,ze^{\beta h}}][1-z\cL_{\beta,h}][1+\cMu_{\beta,ze^{-\beta h}}]=
[1-\cMd_{\beta,ze^{\beta h}}\cMu_{\beta,ze^{-\beta h}}]
\eeq
and
\beq
\label{rid}
[1-z\cL_{\beta,h}]=[1- z\cLd_\beta][1-\cMd_{\beta,ze^{\beta h}}\cMu_{\beta,ze^{-\beta h}}][1- z\cLu_\beta]\;.
\eeq
\end{lem}

\proof
Noting that $[1+ \cMd_{\beta,ze^{\beta h}}]=[1- z\cLd_\beta]^{-1}$ and $[1+\cMu_{\beta,ze^{-\beta h}}]=
[1- z\cLu_\beta]^{-1}$, we calculate directly 
\begin{align}
[1&+ \cMd_{\beta,ze^{\beta h}}][1-z\cL_{\beta,h}][1+\cMu_{\beta,ze^{-\beta h}}]-1\\
&=[1- z\cLd_\beta]^{-1}[1-z\cL_{\beta,h}][1- z\cLu_\beta]^{-1}-1\\
&=[1- z\cLd_\beta]^{-1}(1-z\cL_{\beta,h}-[1- z\cLd_\beta][1- z\cLu_\beta])[1- z\cLu_\beta]^{-1}\\
&=[1- z\cLd_\beta]^{-1}[-z\cLd_\beta z\cLu_\beta][1- z\cLu_\beta]^{-1}\\
&=-\cMd_{\beta,ze^{\beta h}}\cMu_{\beta,ze^{-\beta h}}\;.
\end{align}
Multiplying (\ref{oid}) by $[1- z\cLd_\beta]$ and $[1- z\cLu_\beta]$ from the left and right,
respectively, (\ref{rid}) follows.
\qed

As above, we have the symmetry
\beq
\cMu_{\beta,\tau}=\cS_\beta\cMd_{\beta,\tau}\cS_\beta\;.
\eeq
It is helpful to take advantage of this symmetry by defining 
$\cM_{\beta,\tau}=\cMd_{\beta,\tau}\cS_\beta$ so that
\beq\label{Msquare}
\cMd_{\beta,ze^{\beta h}}\cMu_{\beta,ze^{-\beta h}}=\cM_{\beta,ze^{\beta 
h}}\cM_{\beta,ze^{-\beta h}}\;.
\eeq
Note that the rhs is a square when $h=0$. Utilising (\ref{Msquare}) and Lemma 
\ref{lem1}, we arrive at the following characterisation of eigenvalues and 
eigenfunctions of $\cL_{\beta,h}$.

\begin{prop}\label{lem2}
Let $z^{-1}\notin\left(\sigma(e^{-\beta h}\cLu_{\beta})\cup\sigma(e^{\beta h}\cLd_{\beta})\right)$.
If $f$ is an eigenfunction of $\cM_{\beta,ze^{\beta h}}\cM_{\beta,ze^{-\beta 
h}}$ with eigenvalue $1$,
then $[1+\cMu_{\beta,ze^{-\beta h}}]f$ is an eigenfunction of $\cL_{\beta,h}$ 
with eigenvalue $\lambda=1/z$.
Conversely, if $g$ is an eigenfunction of $\cL_{\beta,h}$ with eigenvalue 
$\lambda=1/z$, then $[1-ze^{\beta h}\cLd_\beta]g$
is an eigenfunction of $\cM_{\beta,ze^{\beta h}}\cM_{\beta,ze^{-\beta h}}$ with 
eigenvalue $1$.
\end{prop}

\proof
If  $f$ is an eigenfunction of $\cM_{\beta,ze^{\beta h}}\cM_{\beta,ze^{-\beta 
h}}$ with eigenvalue $1$, 
then by (\ref{Msquare}) the rhs of (\ref{oid}) acting on $f$ is 
$[1-\cMd_{\beta,ze^{\beta h}}\cMu_{\beta,ze^{-\beta h}}]f=0$. 
Due to the assumption on $z$, the kernels of both $[1+ \cMd_{\beta,ze^{\beta 
h}}]=[1-ze^{\beta h}\cLd_{\beta}]^{-1}$ and 
$[1+\cMu_{\beta,ze^{-\beta h}}]=[1-ze^{-\beta h}\cLu_{\beta}]^{-1}$ are zero, 
so it follows from (\ref{oid}) that $[1-z\cL_{\beta,h}]g=0$ with
$g=[1+\cMu_{\beta,ze^{-\beta h}}]f\neq0$.
The second assertion follows similarly using (\ref{rid}).
\qed

Proposition \ref{lem2} motivates the definition of the set
\beq
\label{def:omega}
\Omega_{\beta,h}=\left\{1/z: 
z\in\mathbb C\setminus\left(\{0\}\cup\sigma(e^{-\beta h}\cLu_\beta)\cup\sigma(e^{\beta h}\cLd_\beta)\right)\mbox{ and }1\in\sigma(\cM_{\beta,ze^{\beta h}}\cM_{\beta,ze^{-\beta h}})\right\}\;.
\eeq
The next proposition relates $\Omega_{\beta,h}$ and $\sigma(\cL_{\beta,h})$.

\begin{prop}\label{propspec}
\beq
\Omega_{\beta,h}=\sigma(\cL_{\beta,h})\setminus
\left(\{0\}\cup\sigma(e^{-\beta h}\cLu_\beta)\cup\sigma(e^{\beta h}\cLd_\beta)\right)\;.
\eeq
\end{prop}

\proof
If $z^{-1}\notin\left(\sigma(e^{-\beta h}\cLu_{\beta})\cup\sigma(e^{\beta h}\cLd_{\beta})\right)$,
then (\ref{rid}) implies that $[1-z\cL_{\beta,h}]^{-1}$ 
exists if and only if $[1-\cMd_{\beta,ze^{\beta h}}\cMu_{\beta,ze^{-\beta h}}]^{-1}$ exists. 
But this is equivalent to the definition of $\Omega_{\beta,h}$, as 
$\Omega_{\beta,h}\cap\left(\sigma(e^{-\beta h}\cLu_\beta)\cup\sigma(e^{\beta h}\cLd_\beta)\right)=\emptyset$.

\qed

Proposition \ref{propspec} will allow us to study the spectral properties of $\cL_{\beta,h}$ by analysing the
spectral properties of $\cM_{\beta,ze^{\beta h}}\cM_{\beta,ze^{-\beta h}}$.

\section{The Function Space}
\label{sect_funct}

We now specify the function space on which $\cL_{\beta,h}$ acts and describe some of its spectral properties on
this space.

Let $\Pi=\all{z\in \mathbb C}{\Re{z}>0}$ denote the  open
right half plane and let $\hip$ denote the space of bounded holomorphic 
functions on $\Pi$. Equipped 
with the norm 
$\norm{f}=\sup_{z\in \Pi}\abs{f(z)}$ the space $\hip$ becomes a Banach
space.

Observe that if $\phi$ is a holomorphic self-map of $\Pi$ and $w\in
\hip$, 
then the operator 
\beq \cC_{w,\phi}:\hip\to \hip \eeq
\beq \cC_{w,\phi}f(z)=w(z)f(\phi(z))\,, \eeq
 known as a \emph{weighted composition operator} 
(see, for example, \cite{cowenbook}), 
is bounded with operator norm $\norm{\cC_{w,\phi}}=\norm{w}$. 
To see this, 
 note that if $f\in \hip$, then $w\cdot f\circ\phi$ is 
holomorphic and bounded on $\Pi$ and 
 \beq \norm{\cC_{w,\phi}}=\sup_{z\in\Pi}\abs{w(z)f(\phi(z))}\leq
 \norm{w}\cdot
\norm{f} \eeq
 Thus $\norm{\cC_{w,\phi}}\leq \norm{w}$. 
But $\norm{\cC_{w,\phi}1}=\norm{w}$, so $\norm{\cC_{w,\phi}}=\norm{w}$ 
as claimed.

Before studying the spectral properties of our operators on $\hip$, 
we require some more notation. We write  
\beq \wu_\beta(z)=\frac{1}{(1+z)^{2\beta}}\,,\quad \phiu(z)=\frac{z}{1+z}\,,\eeq
\beq  \wdd_\beta(z)=1\,,\quad \phid(z)=(1+z)\,,\eeq
so that 
\beq \cLu_\beta=\cC_{\wu_\beta,\phiu} \quad \mbox{and} \quad \cLd_\beta=\cC_{\wdd_\beta,\phid}\,. \eeq 

We shall now consider the spectral properties of our operators in more detail. 
\begin{prop}\label{lem3}\mbox{}\\[-5ex]
\begin{itemize}
\item[(i)] $\cLd_\beta$ is a bounded operator on $\hip$. Its spectrum is the interval $[0,1]$ with every spectral point being an eigenvalue. 
\item[(ii)] $\cLu_\beta$ is a bounded operator on $\hip$ provided that $\Re\beta \geq 0$. 
If $\beta \geq 0$ then $\norm{{\cLu_\beta}^n}=1$ for any $n\in \mathbb N$.
\end{itemize}
\end{prop}

\proof
For the proof of (i) observe that $\phid$ is a holomorphic self-map of $\Pi$ and $\wdd_\beta\in\hip$ so $\cLd_\beta$ is bounded by the argument 
outlined above. The remaining assertions are proved in \cite{gul}. 

We now turn to the proof of (ii). Again, since  $\phiu$ is a holomorphic self-map of $\Pi$ and 
$\wu_\beta \in \hip$ for $\Re\beta \geq 0$ the operator  
$\cLu_\beta$ is bounded by the argument outlined
above.  Suppose now that $\beta \geq 0$. For the norm calculation of
$\cLu_\beta$ observe that $\norm{\wu_\beta}=1$. Thus 
$\norm{\cLu_\beta}=1$ and it follows that 
$\norm{{\cLu_\beta}^n}\leq 1$ for $n\in \mathbb N$. It remains to show that  
$\norm{{\cLu_\beta}^n}\geq 1$. In order to see this, note that if 
$f\in \hip$ is holomorphic at $0$ with $f(0)=1$, then 
 $\cLu_\beta f$ is also holomorphic at $0$ with $\cLu_\beta f(0)=1$. 
Thus ${\cLu_\beta}^n1(0)=1$ for any $n\in \mathbb N$, and so 
 $\norm{{\cLu_\beta}^n}\geq 1$  as claimed. 
\qed

An immediate consequence of the above is the following.

\begin{corollary}
If $\beta \geq 0$ the spectral radii of $\cLd_\beta$ and $\cLu_\beta$ are given by  
\beq r(\cLd_\beta)=r(\cLu_\beta)=1\,. \eeq
\end{corollary}

The following result will play a crucial role in the study of the 
spectral properties of $\cL_{\beta,h}$. 
\begin{prop}\label{prop:compact}
If $\Re \beta\geq 0$ then $\cLd_\beta \cLu_\beta$ is compact. 
\end{prop}

\proof
It is not difficult to see that $\phiu\circ\phid(\Pi)=\all{z\in
  \mathbb C}{\abs{z-\frac34}<\frac14}$. 
Thus $\phiu\circ\phid$ maps all of $\Pi$ into a compact subset of
$\Pi$, and Montel's Theorem \cite[Chapter 1, Proposition 6]{Montel}
implies that $\cC_{1,\phiu\circ \phid}$ is a compact operator on $\hip$. 
But 
\beq \label{prop:compact:eq}
\cLd_\beta \cLu_\beta=\cC_{\wdd_\beta,\phid}\cC_{\wu_\beta,\phiu}=
\cC_{\wu_\beta\circ\phid,\phiu\circ\phid}=
\cC_{\wu_\beta\circ \phid,{\rm id}}\cC_{1,\phiu\circ \phid}\,, 
\eeq
and, since $\wu_\beta\circ \phid\in \hip$, the operator 
$\cC_{\wu_\beta\circ \phid,{\rm id}}$ is bounded, so 
$\cLd_\beta \cLu_\beta$, being the product of a bounded and a compact operator,
 is itself compact. 
\qed

An immediate consequence of the previous proposition is the following
estimate for 
the \textit{essential spectral radius} of 
$\cL_{\beta,h}$ (see, for example,
\cite[Chap. I.4]{edmundsevansbook}).
\begin{prop}\label{lem_ess}
Let $\Re \beta \geq 0$ and $h \in \mathbb C$. The 
essential spectral radius 
of the operator 
$\cL_{\beta,h}=e^{-\beta h}\cLu_\beta+e^{\beta h}\cLd_\beta$ acting on $\hip$
is bounded above by $e^{|\Re(\beta h)|}$.
\end{prop}

\proof Recall that the essential spectral 
radius of an operator $\cA$ can be computed as follows (see, for
example, \cite[Chap. I, Thm. 4.10]{edmundsevansbook}) 
\beq r_{ess}(\cA)=\lim_{n\to\infty}
\left(\inf_{\text{$\cK$ compact}}||\cA^n-\cK||\right)^{1/n}\,.\eeq

Expanding the $n$-th power of $\cL_{\beta,h}$, we find
\beq
\cL_{\beta,h}^n=e^{-n\beta h}{\cLu_\beta}^n+e^{n\beta h}{\cLd_\beta}^n+\cK_n
\eeq
where $\cK_n$ is a sum of $2^n-2$ compact operators. In order to see that they are
compact, note that each them is a product involving the compact operator 
$\cLd_\beta \cLu_\beta$ and bounded operators of the form ${\cLd_\beta}^k$ and ${\cLu_\beta}^l$.
Thus
\beq r_{ess}(\cL_{\beta,h})\leq\limsup_{n\to\infty}||e^{-n\beta h}{\cLu_\beta}^n+e^{n\beta h}{\cLd_\beta}^n||^{1/n} \eeq
and hence, by Proposition~\ref{lem3}, 
\beq r_{ess}(\cL_{\beta,h})\leq
\max\left\{\abs{e^{-\beta h}}r(\cLu_\beta),\abs{e^{\beta h}}r(\cLd_\beta)\right\}
=e^{|\Re(\beta h)|}\,.\eeq
\qed

It turns out that the operator $\cL_{\beta,h}$ has 
a number of 
interesting spectral properties if $\beta\geq 0$ and $h\in \mathbb
R$, because, in this case, $\cL_{\beta,h}$ is a \textit{positive} operator (see
below). Exploiting this additional structure requires some more
terminology, which we now review (for more background see \cite{krasno} or
\cite{krasnoetal}). 

Let $H^\infty_{\mathbb R}(\Pi)=\all{f\in \hip}{f(z)\in \mathbb R
  \mbox{ for $z>0$}}$.  This is a real Banach space when 
equipped with the norm inherited from $\hip$ and its canonical 
complexification equals 
$\hip$ (see \cite[Lemma 5.2]{weights}). 
In $H^\infty_{\mathbb R}(\Pi)$ consider the 
cone $K=\all{f\in H^\infty_{\mathbb R}(\Pi)}{f(z)\geq 0 \mbox{ for
  $z>0$}}$. 
Notice that $K$ is closed and reproducing, 
that is, 
$K-K=H^\infty_{\mathbb R}(\Pi)$. 
For $f,g\in H^\infty_{\mathbb R}(\Pi)$, 
we write $f\leq g$ to mean that $g-f\in K$, 
and this defines a partial order on 
$H^\infty_{\mathbb R}(\Pi)$.  
An operator $\cL$ on $H^\infty_{\mathbb R}(\Pi)$ is said to be 
\emph{positive} (with respect to the partial order induced by $K$) if 
$f\geq 0$ implies $\cL f\geq 0$, or equivalently, 
if $\cL$ leaves $K$ invariant. 

\begin{lem}\label{lem:positive}
Let $\beta\geq 0$ and $h\in \mathbb R$. Then the 
operators $\cLd_\beta$, $\cLu_\beta$, 
and $\cL_{\beta,h}$ are positive 
on $H^\infty_{\mathbb R}(\Pi)$ with respect to $K$.  
\end{lem}

\proof 
Observe that $\wu_\beta\in K$. 
Thus, $f\in K$ implies $\cLu_\beta f\in K$ since 
$\wu_\beta(z)f(\phiu(z))\geq 0$ for $z>0$. Hence $\cLu_\beta$ is positive.   
By a similar argument, $\cLd_\beta$ is positive. 
Finally, $\cL_{\beta,h}$ is positive since it is a sum of positive 
operators.  
\qed 

A consequence of positivity is the following lower bound for the
spectral radius of $\cL_{\beta,h}$. 

\begin{prop}\label{lower}
Let $\beta\geq 0$ and $h \in \mathbb R$. Then the 
spectral radius of the operator 
$\cL_{\beta,h}=e^{-\beta h}\cLu_\beta+e^{\beta h}\cLd_\beta$ acting on $\hip $
is bounded below by $e^{|\beta h|}$.
\end{prop}

\proof
We start with the case $h\geq 0$. Since 
\beq \cL_{\beta,h}1=e^{-\beta h}\wu_\beta+e^{\beta h}1\geq e^{\beta h}1\,, \eeq
the bound $r(\cL_{\beta,h})\geq e^{\beta h}$ follows by 
\cite[Lemma 9.1]{krasnoetal}, 
and the positivity of $\cL_{\beta,h}$. 

Let now $h\leq 0$. For $t>0$ define $f_t(z):=(z+t)^{-2\beta}$. Note
that $f_t\in K$ for any $t>0$. We now claim that for any $t>0$  
\beq \label{lower:hlesseq}
\cL_{\beta,h}f_t\geq \frac{e^{-\beta h}}{(1+t)^{2\beta}}f_t\,.
\eeq 
In order to see this note that for $z>0$ 
\beq 
\cL_{\beta,h}f_t(z) \geq e^{-\beta h}\cLu_\beta f_t(z) 
 = \frac{e^{-\beta h}}{(z+t(1+z))^{2\beta}} 
\geq  \frac{e^{-\beta h}}{(1+t)^{2\beta}}\frac{1}{(1+z)^{2\beta}}
\eeq 
where the last inequality follows since $(z+t(1+z))\leq
(1+t)(1+z)$ for $t,z>0$. 
Now, as before, (\ref{lower:hlesseq}) and the positivity of
$\cL_{\beta,h}$ imply 
\beq  r(\cL_{\beta,h})\geq \frac{e^{-\beta h}}{(1+t)^{2\beta}} \eeq
by \cite[Lemma 9.1]{krasnoetal}, which in turn yields 
$r(\cL_{\beta,h})\geq e^{-\beta h}$ by letting $t\to 0$. 
\qed

We now summarise what we know about the spectral properties of 
$\cL_{\beta,h}$.

\begin{theorem}\label{thm1}
Let $\beta \geq 0$ and $h\in\mathbb R$. 
For the operator $\cL_{\beta,h}$ acting on $\hip$ we have the bounds 
\beq
r_{ess}(\cL_{\beta,h})\leq e^{|\beta h|}\leq r(\cL_{\beta,h})\;.
\eeq
The spectrum of $\cL_{\beta,h}$ in the 
annulus $\all{z\in\mathbb C}{\abs{z}>e^{\abs{\beta h}}}$ coincides with 
$\Omega_{\beta,h}$ and 
consists of isolated eigenvalues of finite algebraic 
multiplicity.
Moreover, if $r(\cL_{\beta,h})>r_{ess}(\cL_{\beta,h})$ then $r(\cL_{\beta,h})$
is an eigenvalue of $\cL_{\beta,h}$.
\end{theorem}

\proof
This follows from Proposition~\ref{lower}, Proposition~\ref{propspec} and
the 
definition of the essential spectral radius. The last assertion follows from Lemma~\ref{lem:positive}
and \cite[Exercise 8.2]{krasnoetal}.
\qed

We end this section with a number of results on the operators  
$\cMd_{\beta,\td}\cMu_{\beta,\tu}$
which will be needed for
the perturbative argument below. A short calculation shows that 
\beq 
\label{mudseries}
\cMd_{\beta,\td}\cMu_{\beta,\tu}f=\sum_{m,n=1}^\infty\td^m\tu^nw_\beta^{(m,n)}\cdot f\circ \phi^{(m,n)}\,, \eeq
with 
\beq w_\beta^{(m,n)}(z)=\frac{1}{(nz+mn+1)^{2\beta}}\quad \mbox{and}\quad \phi^{(m,n)}(z)=\frac{z+m}{nz+mn+1}\,. \eeq

\begin{prop}
\label{holo}
Let 
\beq D_1=\all{(\beta,\td,\tu)
\in {\mathbb C}^3}{\Re\beta >0, \abs{\td}<1,\abs{\tu}<1}\,, \eeq
\beq D_2=\all{(\beta,\td,\tu)
\in {\mathbb C}^3}{\Re\beta >\frac12, \abs{\td}\leq 1,\abs{\tu}\leq 1}\,. \eeq
The function $(\beta,\td,\tu)\mapsto \cMd_{\beta,\td}\cMu_{\beta,\tu}$ 
has the following properties:
\begin{itemize}
\item[(i)] on $D_1$ it is holomorphic in the operator norm topology;
\item[(ii)] on $D_2$ it is continuous in the operator norm topology;
\item[(iii)] on $D_1\cup D_2$ its values are compact operators. 
\end{itemize}
\end{prop}

\proof
First observe that for $\Re \beta \geq 0$ and $z\in \Pi$ 
\beq \abs{w_\beta^{(m,n)}(z) } \le \frac{e^{\pi \abs{\Im \beta }}}{\abs{nz+mn+1}^{2\Re \beta}} \,,\eeq
so 
\beq \norm{w_\beta^{(m,n)}}\leq \frac{e^{\pi \abs{\Im \beta }}}{(mn)^{2\Re\beta}}\,, \eeq
hence
\beq \norm{\cMd_{\beta,\td}\cMu_{\beta,\tu}}\leq \sum_{m,n=1}^\infty\abs{\td}^m\abs{\tu}^n  \frac{e^{\pi \abs{\Im \beta }}}{(mn)^{2\Re\beta}} \,, \eeq
which means that the series (\ref{mudseries}) converges in the operator norm topology for any $(\beta,\td,\tu)\in D_1\cup D_2$. 

Assertions (i) and (ii) now follow by observing that $\beta \mapsto w_\beta^{(m,n)}$ is holomorphic (and thus continuous) for $\Re\beta >0$ in the norm topology 
on $\hip$ for every $m,n\in \mathbb N$. For the proof of (iii) we note that for any fixed $\Re\beta>0$,  $\abs{\td},\abs{\tu}<1$ 
\beq \cMd_{\beta,\td}\cMu_{\beta,\tu}=\td\tu[1-\td\cLd_\beta]^{-1}\cLd_\beta \cLu_\beta [1-\tu\cLu_\beta]^{-1}\,, \eeq
so $\cMd_{\beta,\td}\cMu_{\beta,\tu}$ is compact by Proposition~\ref{prop:compact}. The remaining assertion now follows from (ii) and the fact that the operator norm limit 
of compact operators is itself a compact operator. 
\qed

\begin{prop}
\label{wasdanachkommt}
For $\beta>\frac12$, $0<\td,\tu\leq 1$, 
the operator $\cMd_{\beta,\td}\cMu_{\beta,\tu}$ has a 
simple leading eigenvalue. 
\end{prop}
\proof Fix $\beta>\frac12$, $0<\td,\tu\leq 1$. 
Observe that $\cMd_{\beta,\td}\cMu_{\beta,\tu}$ is a transfer operator corresponding to a real analytic full branch expanding map on $[0,1]$ with strictly 
positive weights. The proof of \cite[Proposition 4.9]{weights} now shows that $\cMd_{\beta,\td}\cMu_{\beta,\tu}$ is $1$-positive with respect to $K$. Since 
$\cMd_{\beta,\td}\cMu_{\beta,\tu}$ is compact by Proposition~\ref{holo} the assertion now follows from \cite[Theorems 2.5, 2.10 and 2.13]{krasno} and the fact that 
the canonical complexification of $H^\infty_{\mathbb R}(\Pi)$ is $\hip$. 
\qed

\section{Dynamical Systems}\label{sect_dyn}

In this section we provide explicit representations of the operators 
defined above and show the connection to the Gauss map. 

The operators $\cMu_{\beta,\tau}$, $\cMd_{\beta,\tau}$, and $\cM_{\beta,\tau}$
have explicit power series expansions in $\tau$, given by
\beq
\cMu_{\beta,\tau}f(x)=\sum_{n=1}^\infty\frac{\tau^n}{(1+nx)^{2 
\beta}}f\left(\frac x{1+nx}\right)
\quad \; \text{,} \quad
\cMd_{\beta,\tau}f(x)=\sum_{n=1}^\infty \tau^nf(x+n)\;,
\eeq
and
\beq
\label{eq:gauss}
\cM_{\beta,\tau}f(x)=\sum_{n=1}^\infty\frac{\tau^n}{(n+x)^{2 
\beta}}f\left(\frac 1{n+x}\right)
\;.
\eeq

\begin{figure}[ht]
\centering\includegraphics[width=10.0cm]{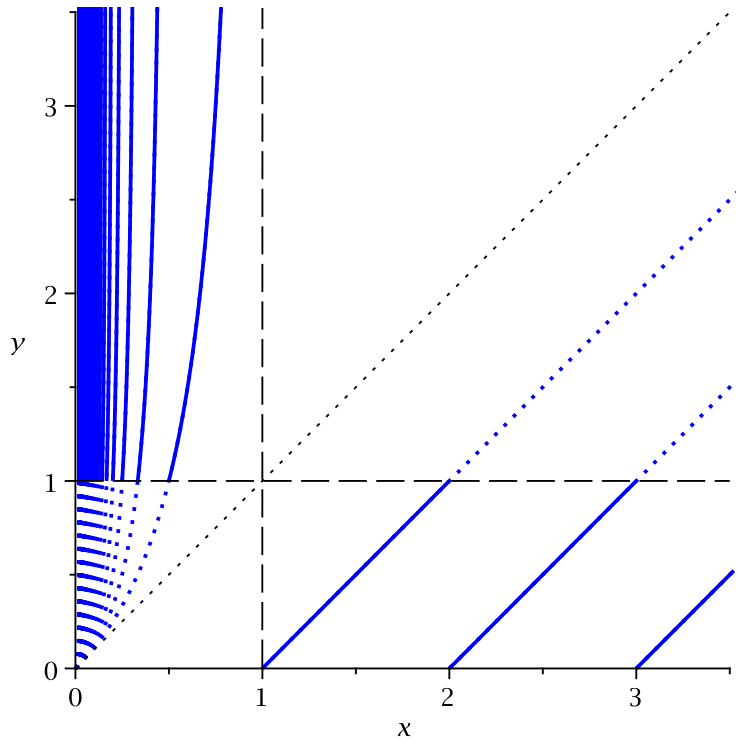}
\caption{The graph of the left-most and right-most branches of the 
iterated transformations $T^n$ for $n\in\mathbb N$. Restricting 
the iterates as indicated by the solid lines, one obtains the 
interval map $y=\hat T(x)$.} \label{fig2}
\end{figure}

In order to find an interpretation of these operators as weighted 
transfer operators associated to interval maps, note that
the operators $\cMu_{\beta,\tau}$ and $\cMd_{\beta,\tau}$ are given by 
the collection of the left-most and right-most branches, respectively, 
of the iterated transformations $T^n$ for $n\in\mathbb N$, where $T$ is given in 
Eqn.~(\ref{trafo}). This is indicated in Fig.~\ref{fig2}.

If one restricts the maps as indicated by the solid lines in 
Fig.~\ref{fig2}, one sees that another interval map $\hat T$ is formed on $\mathbb R^+$.
Clearly the map $\hat T$ exchanges the intervals $(0,1)$ and $(1,\infty)$.

\begin{figure}[ht]
\centering\includegraphics[width=10.0cm]{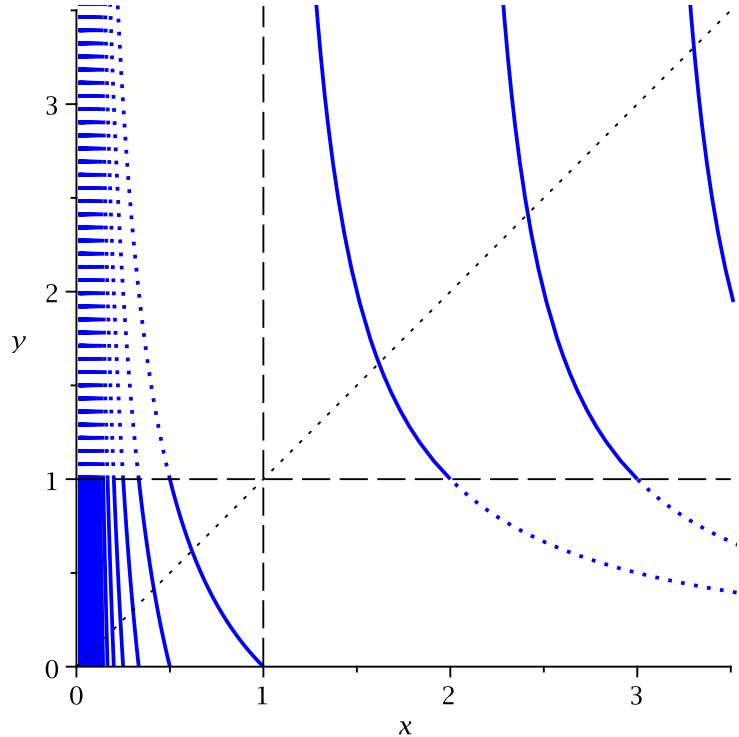}
\caption{The graph of $y=\hat TS(x)$ (solid lines), along with the analytic extension
of its branches (dotted lines).} \label{fig3}
\end{figure}

Recall that the operator $\cM_{\beta,\tau}$ was defined by composition with
$\cS_\beta$ as
\beq
\cM_{\beta,\tau}=\cMd_{\beta,\tau}\cS_\beta=\cS_\beta\cMu_{\beta,\tau}\;.
\eeq
The transformation underlying the operator $\cM_{\beta,\tau}$
is obtained from Fig.~\ref{fig2} by composing $\hat T$ with $S(x)=1/x$. It is easy to see that
$\hat TS=S\hat T$. The graph of $\hat TS$ is shown in Fig. \ref{fig3}. The associated
dynamical system is split into two independent subsystems on the intervals $(0,1)$ and $(1,\infty)$, respectively.

Restricting $\cM_{\beta,1}$ to act on functions on the unit interval $[0,1]$ 
gives precisely the Ruelle-Perron-Frobenius transfer operator of the Gauss map 
$x \mapsto 1/x \mod 1$, 
which has branches
\beq \label{branch}
\hat T_{n}(x)=\frac{1}{x}-n \;.
\eeq
Accordingly, $\cM_{\beta,\tau}$ may be regarded as a generalised transfer operator of the Gauss map, with the branches  weighted differently.  

This interpretation will be a key ingredient in the perturbative expansion in
the next section.

\section{Perturbation Theory} \label{sec3}

In this section we come to the final point of our analysis.  We employ Proposition 
\ref{lem2}, which connects eigenvalues of $\cM_{\beta,ze^{\beta h}}\cM_{\beta,ze^{-\beta h}}$ to eigenvalues of $\cL_{\beta,h}$ to make a perturbation expansion around the critical point $(\beta,h)=(1,0)$. More explicitly, for $\beta$ in a left neighbourhood
of $1$, we will choose $z$ and $h$ so that $\cM_{\beta,ze^{\beta h}}\cM_{\beta,ze^{-\beta h}}$ 
has eigenvalue $1$. This results in an implicit equation for the inverse eigenvalue $z(\beta,h)$ which by (\ref{flam}) leads to the asymptotic form of the free energy $f(\beta,h)$.

The central object in the perturbative calculation is the operator product
\beq \label{cPdef}
 \cP_{\beta,ze^{\beta h},ze^{-\beta h}}=
 \cM_{\beta,ze^{\beta h}}\cM_{\beta,ze^{-\beta h}}\;. 
\eeq
By Propositions \ref{holo} and \ref{wasdanachkommt}, we know that near the critical point, i.e. $z=1$, $\beta=1$, and $h=0$,
this is a holomorphic compact operator-valued function that extends continuously to this point. Furthermore, at the critical point,
$\cP_{1,1,1}$ has a simple leading eigenvalue.

To keep notational overload to a minimum, we write
\beq
\vec\tau=(\td,\tu)=(ze^{\beta h},ze^{-\beta h})
\eeq
and put $\cP_{\beta,{\vec \tau}}=\cP_{\beta,ze^{\beta h},ze^{-\beta h}}=\cM_{\beta,ze^{\beta h}}\cM_{\beta,ze^{-\beta h}}$. 
We further omit any of the variables ($\beta,\td,\tu$) when they take on their
respective value at the critical point, i.e. $z=1$, $\beta=1$, or $h=0$.
For example, we write $\cP=\cP_{1,1,1}$, $\cP_{\vec\tau}=\cP_{1,ze^h,ze^{-h}}$
and so on. We use similar conventions for other quantities.

Now we find formally that
\beq \label{invdens}
\cL h=h \quad \text{with} \quad h(x)=\frac1x \;,
\eeq
however the function $h$ is not bounded, and therefore not an eigenfunction of the operator in the space $\hip$.

On the other hand, the corresponding equation for $\cP=\cP_{1,1,1}$ is 
\beq\label{stuff}
\cP g=g\quad\text{with}\quad g(x)=\frac1{\log(2)}\frac1{1+x}\;,
\eeq
and we can check easily that indeed formally $h=\log(2)(1+\cMu_{1,1})g$ as expected. 
The function $g$ lies in $\hip$, and is therefore an eigenfunction of $\cP$.

As indicated in the previous section, $\cP$ is the Perron-Frobenius operator of
the second iterate of the Gauss map on the unit interval. It follows that the 
left eigenfunction of $\cP$ is $\mu=\mu_L$, the Lebesgue measure on $[0,1]$.

Integration of $1/(1+x)$ with respect to the Lebesgue measure on $[0,1]$ gives 
$\log(2)$, which motivates the normalisation of $g$.

Given the eigenvalue equations
\beq
\cP_{\beta,\vec\tau}g_{\beta,\vec\tau}=\lambda_{\beta,\vec\tau}g_{\beta,\vec\tau}\;,
\quad
\mu_{\beta,\vec\tau}\cP_{\beta,\vec\tau}=\lambda_{\beta,\vec\tau}\mu_{\beta,\vec\tau}\;,
\eeq
we shall now solve the equation
\beq
\label{goal}
\lambda_{\beta,\vec\tau}=1
\eeq
perturbatively around $\beta=1$ and $\vec\tau=(1,1)$, proceeding as in \cite{P1}.

By Proposition \ref{holo}, the compact operator $\cP_{\beta,\vec\tau}$ is an analytic function of 
$\beta$ for $\Re\beta>0$ in the operator norm topology. Thus, we can
expand $\cP_{\beta,\vec\tau}$ around $\beta=1$ as
\beq
\label{eq:expans}
\cP_{\beta,\vec\tau}=\cP_{\vec\tau}+\sum_{n=1}^\infty(1-\beta)^n\cP_{\vec\tau}^{(n)}\;.
\eeq
Moreover, since by Proposition~\ref{wasdanachkommt} the leading eigenvalue $\lambda_{\beta,\vec\tau}$ is simple, therefore it 
  is analytic in $\beta$ and continuous in $\vec\tau$. By the
same argument, $g_{\beta,\vec\tau}$ ($\mu_{\beta,\vec\tau}$) is holomorphic in $\beta$
and continuous in $\vec\tau$ with respect to the norm topology on $\hip$
(the strong dual topology on the dual of $\hip$).
We thus have expansions analogous to (\ref{eq:expans}) for the eigenvalues $\lambda_{\beta,\vec\tau}$
and the left and right eigenfunctions $g_{\beta,\vec\tau}$ and $\mu_{\beta,\vec\tau}$,
respectively.

We choose the normalisation $\mu_{\beta,\vec\tau}g_{\beta,\vec\tau}=1$. Expanding 
\beq
\mu_{\vec\tau}\cP_{\beta,\vec\tau}g_{\beta,\vec\tau}=\lambda_{\beta,\vec\tau}
\mu_{\vec\tau}g_{\beta,\vec\tau}
\eeq
and
\beq
\mu_{\beta,\vec\tau}\cP_{\beta,\vec\tau}g_{\vec\tau}=\lambda_{\beta,\vec\tau}\mu_{\beta,\vec\tau}g_{\vec\tau}
\eeq
to lowest orders in $(1-\beta)$ and comparing coefficients, we find for the first-order 
change of the eigenvalue
\beq
\lambda_{\vec\tau}^{(1)}=\mu_{\vec\tau}\cP_{\vec\tau}^{(1)}g_{\vec\tau}\;,
\eeq
which is a standard result of first-order perturbation theory \cite{kato}. We therefore have
\begin{align}
\lambda_{\beta,\vec\tau}&=\lambda_{\vec\tau}+
(1-\beta)\mu_{\vec\tau}\cP_{\vec\tau}^{(1)}g_{\vec\tau}+O((1-\beta)^2)\\
&=\lambda_{\vec\tau}+(1-\beta)\mu\cP^{(1)}g[1+o(1-\tu)+o(1-\td)]+O((1-\beta)^2)\;,
\label{estimate}
\end{align}
where for the final estimate we have used continuity in $\vec\tau$.

From $\cP_\beta=\cM_\beta^2$, where $\cM_{\beta}=\cM_{\beta,1}$ is the 
transfer operator for the Gauss map, it follows that
\beq
\mu\cP^{(1)}g=\mu\cM^{(1)}\cM g+\mu\cM\cM^{(1)}g=2\mu\cM^{(1)}g\;,
\eeq
where we have expanded $\cM_\beta=\cM+(1-\beta)\cM^{(1)}+O((1-\beta)^2)$.

By a standard result \cite{beck93}, this can be expressed in terms of the Lyapunov exponent of the associated interval map.
Here, one obtains (see e.g. \cite{pollicott}) the Lyapunov exponent $\lambda_G$ of the Gauss map,
\beq
\lambda_G=\mu\cM^{(1)}g=-\left.\frac\partial{\partial\beta}\right|_{\beta=1}\mu\cM_{\beta}g=\frac{\pi^2}{6\log(2)}\;.
\eeq
Therefore Eqn.~(\ref{estimate}) gives
\beq \label{lambtfinal}
\lambda_{\beta,\vec\tau}=\lambda_{{\vec \tau}}+2\lambda_G(1-\beta)[1+o(1-\tu)+o(1-\td)]+O((1-\beta)^2)\;.
\eeq

Next, we consider the ${\vec\tau}$ dependence of  $\lambda_{\vec\tau}=\lambda_{1,{\vec\tau}}$. 
Using $\lambda_{{\vec \tau}}\mu g_{\vec \tau}=\mu \cP_{{\vec \tau}}g_{\vec \tau}$,
we rewrite
\beq
\mu\cP_{{\vec \tau}}\,g_{\vec \tau}-\mu\cP \,g_{\vec 
\tau}=(\lambda_{\vec \tau}-1)\mu g_{\vec \tau}\;.
\eeq
Hence
\beq \label{lamt}
\lambda_{\vec \tau}=1+\frac{\mu(\cP_{{\vec \tau}}-\cP) g_{\vec 
\tau}}{\mu(g_{\vec \tau})}
=1+\frac{\mu(\cP_{{\vec \tau}}-\cP) g+\mu(\cP_{{\vec \tau}}-\cP) (g_{\vec 
\tau}-g)}{1+\mu(g_{\vec \tau}-g)}\;,
\eeq 
where we have used the normalisation condition $\mu g =1$. Continuity in $\vec\tau$
implies that
\beq \label{lamt2}
\lambda_{\vec \tau}=1+\mu(\cP_{{\vec \tau}}-\cP) g[1+o(1-\tu)+o(1-\td)]\;.
\eeq 
Combining Eqns.~(\ref{goal}), (\ref{lambtfinal}), and (\ref{lamt2}), this implies
\beq
\label{foo}
-\mu(\cP_{{\vec \tau}}-\cP) g\sim2\lambda_G(1-\beta)
\eeq
as $\beta\to1$ (and, hence, both $\tu\to1^-$ and $\td\to1^-$).

The final step lies in the estimate of $\cP_{\vec \tau}-\cP$. We write
\beq
\cP_{\vec \tau}-\cP=\cM_{\td} 
\cM_{\tu}-\cM^2=(\cM_{\td}-\cM)\cM+\cM(\cM_{\tu}-\cM)+(\cM_{\td}-\cM)(\cM_{\tu}-
\cM)\;.
\eeq

Eqn.~(\ref{eq:gauss}) implies that
\beq
||\cM_\tau-\cM|| \leq \eta(\tau)\;.
\eeq
Here $\eta(\tau) = \sum_{n=1}^{\infty}\frac{1-\tau^n}{n^2}=\Li_2(1)-\Li_2(\tau)$, 
where $\Li_2$ denotes the dilogarithm.  
It follows immediately that $||(\cM_{\td}-\cM)(\cM_{\tu}-\cM)|| 
\le \eta (\td)\eta (\tu)$.  Hence, 
\beq
\cP_{\vec \tau} - \cP = 
(\cM_{\td}-\cM) \cM + \cM (\cM_{\tu}-\cM)+ O(\eta (\td)\eta (\tu)) 
\eeq
in operator norm.  Hence, by (\ref{foo}),
\beq \label{foofinal}
-\mu(\cM_{\td}-\cM)g+\mu(\cM_{\tu}-\cM)g\sim2\lambda_G(1-\beta) \;.
\eeq
An explicit  calculation then gives the exact expression
\beq \label{mMg}
\mu(\cM_{\tau}-\cM)g = -\frac{(1-\tau)^2}{\tau^2 \log(2)} \sum_{n=1}^{\infty} 
\tau^n \log n.
\eeq
The asymptotic form follows on writing 
\beq
\log n=\sum_{k=1}^n\frac1k-\gamma-\frac1{2n}+O(n^{-2})\;,
\eeq
where $\gamma=0.5772\ldots$ is the Euler-Mascheroni constant. Inserting
this in (\ref{mMg}) gives immediately
\beq
\sum_{n=1}^{\infty} \tau^n 
\log n=\frac1{1-\tau}\log\frac1{1-\tau}-\gamma\frac1{1-\tau}
       -\frac12\log\frac1{1-\tau}+O(1)\;,
\eeq
uniformly for $|\tau|<1$.

We are now in a position to obtain the asymptotic expansion of the free 
energy. Inserting the leading order asymptotics of $\mu(\cM_{\tau}-\cM)g$
into equation (\ref{foofinal}), we arrive at
\beq
[-(1-\td)\log(1-\td)-(1-\tu)\log(1-\tu)]\sim
2\log(2)\,\lambda_G(1-\beta) \;.
\eeq
Substituting $\td=e^{\beta(f+h)}$ and $\tu=e^{\beta(f-h)}$ and expanding for 
small $f$ and $h$ gives to leading order
\beq
\label{result}
2\log(2)\,\lambda_G (1-\beta) \sim (f+h)\log(-(f+h))+(f-h)\log(-(f-h)) \;.
\eeq
Setting $\beta_c=1$ and $C=\log(2)\,\lambda_G=\pi^2/6$, we see that this is, aside from 
constants, the same as equation (40) in \cite{PFK}, which was found using a 
cluster approximation.  The analysis therein then immediately gives 
(cf.~equation (46) in \cite{PFK})
\beq \label{fres}
f \sim \frac t{\log t}-\frac1{2} \frac{h^2}t \quad \text{for} \quad h^2\ll t 
\ll 1 \;,
\eeq
 where the rescaled temperature variable $t$ is given by $t=2\log(2)\lambda_{G}(1-\beta)$.
Therefore the temperature deviation from the critical point is scaled by the Lyapunov 
exponent of the Gauss map, 
$\lambda_G=\frac{\zeta(2)}{\log(2)}=\frac{\pi^2}{6\log(2)}$. Note that, in 
addition, (\ref{fres}) implies that $\lambda_G$ determines the amplitude of 
both the specific heat $C$ and susceptibility $\chi$ singularities, with $C$
proportional to $1/\lambda_G$ and $\chi$ proportional to $\lambda_G$.

Also, just as in \cite{PFK}, the asymptotic shape of the phase boundary is given by
letting $-f=|h|=h_c$ in (\ref{result}). We obtain for the dependence of the critical field
strength $h_c$ to leading order
\beq
h_c\sim\frac t{\log t}\;.
\eeq

\section{Discussion} \label{sec4}

It is of interest to discuss the relation of the present treatment to the cluster
approximation presented in \cite{PFK}. In that approximation the central quantity
is the cluster generating function $\Lambda(\beta,\tau)$. Here we do not linearise the
dynamical map, with the result that the function $\Lambda(\beta,\tau)$ is replaced by
the operator $\cM_{\beta,\tau}$. Conversely, the effect of linearising the map on 
the operator is that it becomes a multiplication operator when acting on constant functions.

Physically, this corresponds to replacing a complicated system with interactions of all types  
(see \cite{KO} and references therein) by non-interacting clusters.
The significance of our work lies in the fact that the behaviour of both models near the
critical point is identical, thus justifying the cluster approximation.  Note that the resulting  
non-interacting cluster model
is similar to the ones discussed in \cite{FF}.

As mentioned, the renormalisation group result for $f(\beta,h)$ found in 
\cite{FK} does not quite agree with (\ref{fres}). Specifically,  the second 
term has the form $ \frac{h^2 \log t }t$, which as $t \to 0$, is larger than 
the corresponding term in (\ref{fres}).  As discussed in \cite{FK}, there does 
not seem to be any consistent way to remove this term in the renormalisation 
group framework.  However, this is perhaps not so surprising, since the Farey 
model is known to have long-range interactions (see \cite{KO} and references therein), 
which renders results from a renormalisation group treatment questionable.

Finally, although  our results are exact, we indicate which points of our treatment are not quite rigorous.  
The results given in section \ref{sec2} are rigorous, but the problem of pointwise evaluation is not 
completely settled, and the spectrum of $\cP_{\beta,\vec\tau}$ has not been fully characterized.
In particular, the possibility of another leading eigenvalue has not be ruled
out.  In addition, our particular choice of a function space might seem unusual in that it does not 
respect the ``spin flip" symmetry of the model.  However, it does not seem possible to
find a function space respecting this symmetry for which the perturbation calculations employed are tractable.

 \section{Acknowledgements}
We thank Gerhard Keller and Hans-Henrik Rugh for helpful discussions, and E. Lieb 
for bringing \cite{LW} to our attention. Thomas Prellberg
thanks LASST and the University of Maine for hospitality. Peter Kleban thanks
the School of Mathematical Sciences, Queen Mary, University of London, for hospitality. This work 
was supported in part by the National Science Foundation Grant No. DMR-0536927, the Libra Foundation, 
and the London Mathematical Society.

\end{document}